\documentclass[conference]{IEEEtran}
\ifCLASSINFOpdf
\else
\fi
\usepackage{pgfplots}
\usepackage[hidelinks,bookmarks=false]{hyperref}
\begin{document}
\begin{sloppy}
\title{NLP2Code: Code Snippet Content Assist\\via Natural Language Tasks}
\author{\IEEEauthorblockN{Brock Angus Campbell and Christoph Treude}
\IEEEauthorblockA{School of Computer Science\\
University of Adelaide\\
Adelaide, SA, Australia\\
a1687816@student.adelaide.edu.au, christoph.treude@adelaide.edu.au}
}
\maketitle
\begin{abstract}
Developers increasingly take to the Internet for code snippets to integrate into their programs. To save developers the time required to switch from their development environments to a web browser in the quest for a suitable code snippet, we introduce NLP2Code, a content assist for code snippets. Unlike related tools, NLP2Code integrates directly into the source code editor and provides developers with a content assist feature to close the vocabulary gap between developers' needs and code snippet meta data. Our preliminary evaluation of NLP2Code shows that the majority of invocations lead to code snippets rated as helpful by users and that the tool is able to support a wide range of tasks.

Video: \url{https://www.youtube.com/watch?v=h-gaVYtCznI}
\end{abstract}
\IEEEpeerreviewmaketitle
\section{Introduction and Motivation}

The Internet has provided software developers with an effective infrastructure for sharing and accessing programming knowledge, often curated through social media mechanisms, such as voting or commenting~\cite{Treude12}. The prime example of this is the Question and Answer website Stack Overflow\footnote{\url{http://stackoverflow.com/}} with more than 14 million questions and more than 22 millions answers as of July 2017. Many of the questions and answers contain code snippets~\cite{Treude11}, and much research has been conducted on the quality of these snippets (e.g.,~\cite{Calefato15, Nasehi12}).

Question and Answer websites are not primarily designed for direct code reuse~\cite{Barzilay13}. With the current tooling available, a common scenario is for a developer to write code in an Integrated Development Environment (IDE), reach a point at which help is needed, switch to a web browser, type a query into a search engine, evaluate search results, find a suitable solution, and integrate the newly acquired knowledge into the source code in the IDE---often in the form of a code snippet.

\begin{figure}
\centering
\includegraphics[width=\linewidth]{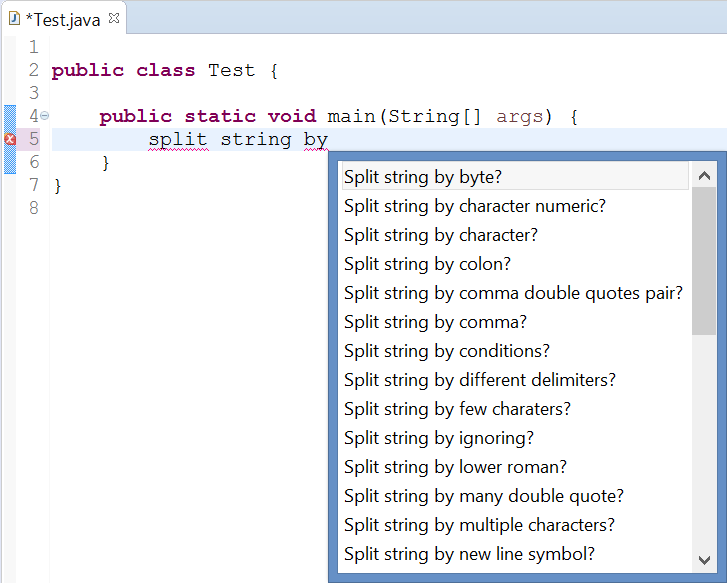}
\caption{NLP2Code's content assist}
\label{fig:autocomplete}
\vspace{-.5cm}
\end{figure}

\begin{figure*}
\centering
\includegraphics[width=\linewidth]{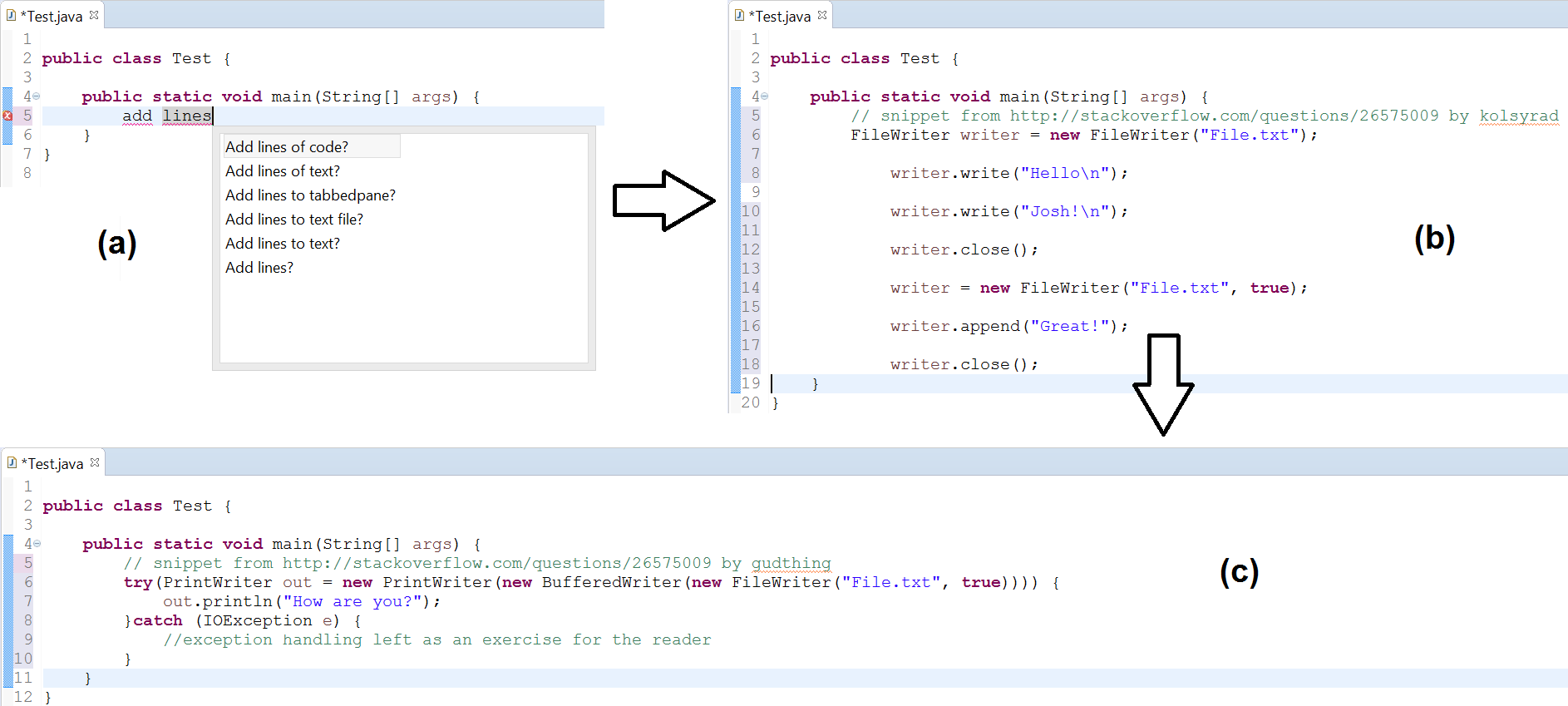}
\caption{NLP2Code's flow. To transition from (a) to (b), select a content assist suggestion, and to cycle through code snippets, such as (c), press \texttt{Ctrl+`}}
\label{fig:flow}
\end{figure*}

Following the argument that too many context switches between different tools can significantly impact the productivity of a developer~\cite{Proksch15}, we introduce NLP2Code, a plugin for the Eclipse IDE, which allows developers to integrate code snippets from Stack Overflow into their code without leaving the source code editor. Unlike previous efforts to integrate Stack Overflow content into the IDE, such as Seahawk~\cite{Ponzanelli13}, Prompter~\cite{Ponzanelli14}, and T2API~\cite{Nguyen16}, NLP2Code does not introduce additional views or windows to the IDE, but works solely in the source code editor.

Similar to Eclipse's code completion mechanism~\cite{Bruch09}, we introduce a content assist for code snippets. This content assist is aimed at closing the vocabulary gap~\cite{Mika09} between developers' needs and relevant Stack Overflow threads, making suggestions based on natural language tasks that we extract from Stack Overflow thread titles, using our previous work on extracting tasks from software documentation~\cite{Treude15a, Treude15b}.

Our preliminary evaluation of NLP2Code with 6 participants and 101 invocations showed that the majority of invocations led to helpful snippets, often based on the first result, and that NLP2Code's responses to these queries outperformed the state of the art. Participants used NLP2Code for a wide range of tasks from looking up implementations of particular algorithms to finding API usage examples and template code.

Compared to previous work, we make three contributions:

\begin{itemize}
    \item a code snippet content assist feature, designed to close the vocabulary gap between developers' needs and code snippet meta data,
    \item seamless integration of code snippets from Stack Overflow into code without having to leave the source code editor, and
    \item a preliminary evaluation with 101 queries, including a comparison to the state of the art.
\end{itemize}

\section{NLP2Code}

We developed NLP2Code as a plugin to the Eclipse IDE. However, none of the underlying concepts are Eclipse or Java-specific, and we only focused on this ecosystem since we had access to undergraduate students working with Eclipse for a preliminary evaluation and since Eclipse offers a plugin infrastructure that has already benefited many other research projects (e.g.,~\cite{Bruch09}).

After installing the NLP2Code plugin, a user can open its content assist anywhere in the source code editor by pressing \texttt{Ctrl+1}.\footnote{All key bindings are configurable.} Figure~\ref{fig:autocomplete} shows a screenshot of the content assist after typing \textit{``split string by''}.\footnote{\textit{``split string by whitespaces''} was one of the tasks that our study participants completed using the NLP2Code content assist.} As shown in this example, even if a user does not exactly know how to specify what they want to split strings by, the content assist helps close this vocabulary gap by providing a large number of suggestions for which NLP2Code can provide code snippets. The content assist suggestions are the 599,550 natural language tasks that the TaskNav algorithm~\cite{Treude15a, Treude15b} extracted from the 1,109,677 Stack Overflow thread titles tagged ``java'' in the latest Stack Overflow data dump.\footnote{\url{https://archive.org/details/stackexchange}, up to 12 tasks per title.} 

The TaskNav algorithm conceptualizes tasks in software documentation as verbs associated with a direct object and/or a prepositional phrase, such as \textit{``get iterator''}, \textit{``get iterator for collection''}, and \textit{``add to collection''}. After removing markup and pre-processing code elements and incomplete sentences (see~\cite{Treude15a} for details), the algorithm makes use of the grammatical dependencies---relations between words in a sentence as identified by the Stanford NLP toolkit~\cite{Manning14}. Because tasks can be described in different grammatical ways (e.g., \textit{``returning an iterator''}, \textit{``return iterator''}, \textit{``iterator returned''}, and \textit{``iterator is returned''}), dependencies between words in active and passive voice are considered. In addition, context might be important, e.g., whether an iterator is returned or whether the documentation instructs the user to \textit{``not return iterator''}. Furthermore, the iterator might be specified using additional words, such as \textit{``list iterator''}, and prepositional phrases might make the task description more specific, such as \textit{``return iterator of collection''}. For example, TaskNav extracted the task \textit{``add lines to text file''} from the Stack Overflow question title \textit{``Best strategy to add lines of text to a text file''}.\footnote{\url{http://stackoverflow.com/q/26575009}} TaskNav uses a handcrafted list of about 200 programming actions and about 30 generic objects to filter out tasks irrelevant to software development.

Our assumption that developers need assistance specifically for tasks is supported by the benchmark used to evaluate \textsc{swim}~\cite{Raghothaman16} where 25 of the 30 benchmark queries follow a similar task pattern. In addition to providing content assist suggestions, the task-based approach helps overcome the limitations of the commonly used bag-of-words assumption and can distinguish between tasks such as \textit{``convert int to string''} and \textit{``convert string to int''}~\cite{Gu16}.

Content assist suggestions in NLP2Code can be filtered by typing characters and words, and pressing \texttt{Enter} selects a suitable suggestion. After the selection, NLP2Code conducts a query using the Google Custom Search Engine\footnote{\url{https://cse.google.com.au/cse/}} on \url{http://stackoverflow.com/}. The plugin then collects up to three code snippets (i.e., content surrounded by \texttt{<pre><code>} tags) from answers with the highest scores from the four first Stack Overflow threads returned by the search. We chose these numbers by trading off performance (impacted by website queries) and diversity of results (by considering different threads). The code snippet from the answer with the highest score in the first thread is automatically inserted into the source code editor along with a comment indicating its source, replacing the original text. The user can then cycle through different code snippets using \texttt{Ctrl+`}.

Figure~\ref{fig:flow} shows an example of this workflow from our preliminary evaluation. In this case, the participant required help with \textit{``add lines to text file''} and NLP2Code assisted with the suggestions shown in Figure~\ref{fig:flow}(a) after the user typed \textit{``add lines''}. Figure~\ref{fig:flow}(b) shows the first code snippet that NLP2Code returned along with the comment indicating its source. In this case, the participant already rated the first code snippet as helpful. Another option would have been to cycle through further code snippets using \texttt{Ctrl+`}. Figure~\ref{fig:flow}(c) shows the second snippet that NLP2Code would have returned.

In addition to the content assist described above, NLP2Code can be invoked by highlighting any text in the Eclipse IDE and pressing \texttt{Ctrl+6} or by surrounding text with question marks (e.g., \textit{``?add lines to text file?''}).

\section{Example Scenarios}

In this section, we discuss two example scenarios for developers using NLP2Code.

\paragraph{API usage}

One scenario for the use of NLP2Code is that of learning how to use an API. As an example, one of our study participants needed help with \textit{``add custom jpanel to jframe''}. For this task, the participant selected the code snippet in Stack Overflow answer 22621494\footnote{\url{http://stackoverflow.com/a/22621494}}, which was the second code snippet that NLP2Code returned. The code snippet consists of ten lines of code, first setting up instances of \texttt{JFrame} and \texttt{JPanel} before adding the latter to the former. This code snippet was rated as helpful by the participant.

\paragraph{Algorithms}

Another group of tasks that our participants completed with NLP2Code is the reuse of algorithms. One of our study participants needed help with \textit{``complete bubble sort''}. The first code snippet returned by NLP2Code came from Stack Overflow answer 16089042\footnote{\url{http://stackoverflow.com/a/16089042}}, and it consists of a method implementing bubble sort in 13 lines of code. The participant rated this snippet as helpful.

\section{Preliminary Evaluation}

\begin{table}
\centering
\caption{Helpful and unhelpful NLP2Code invocations}
\label{tab:queries}
\begin{tabular}{lrr}
\hline
                    & NLP2Code & T2API \\
\hline
correct/helpful     & 74       & 16    \\
incorrect/unhelpful & 27       & 45    \\
no code snippet     & 0        & 40    \\
\hline
\end{tabular}
\end{table}

\begin{figure}
\centering
\begin{tikzpicture}
\begin{axis}[
    y = 0.05cm,
    ylabel = invocations,
    xlabel = ``next snippet'' calls,
    symbolic x coords={0, 1, 2, 3, 4, 5, 6, 7, 8, 9, 10},
    xtick=data,
    ymin = 0]
    \addplot[ybar,fill=white] coordinates {
    (0, 52)
    (1, 17)
    (2, 12)
    (3, 7)
    (4, 2)
    (5, 3)
    (6, 4)
    (7, 1)
    (8, 1)
    (9, 1)
    (10, 1)
    };
    \addplot[ybar,fill=black] coordinates {
    (0, 43)
    (1, 13)
    (2, 9)
    (3, 5)
    (4, 1)
    (5, 1)
    (6, 0)
    (7, 1)
    (8, 0)
    (9, 0)
    (10, 1)
    };
\end{axis}
\end{tikzpicture}
\caption{Number of ``next snippet'' calls; black denotes helpful snippets and white denotes unhelpful snippets}
\label{fig:cycles}
\end{figure}
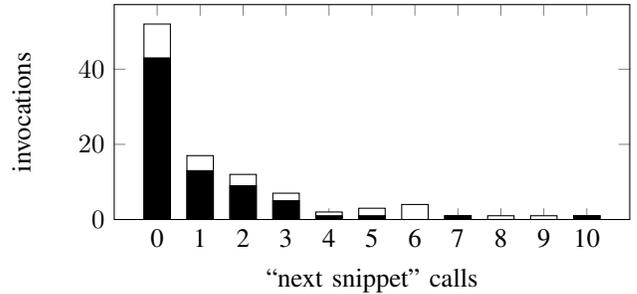

\begin{table}
\centering
\caption{Invocations with content assist (excerpt)}
\label{tab:examples-auto-complete}
\begin{tabular}{l}
\hline
implement dijkstra algorithm      \\
complete bubble sort              \\
sort array for binary search      \\
convert uppercase to lowercase    \\
add custom jpanel to jframe       \\
call sleep on current thread      \\
find number of shortest paths     \\
implement depth first search      \\
compile basic hello world program \\
generate random integers          \\
\hline
\end{tabular}
\end{table}

\begin{table}
\centering
\caption{Invocations without content assist (excerpt)}
\label{tab:examples-without}
\begin{tabular}{l}
\hline
initialize two dimensional vector                \\
round double value to two decimal places        \\
download html from a website                    \\
remove html tags from a string                  \\
evaluate math expression in string form        \\
play a sound                                    \\
calculate difference between two date instances \\
generate all permutations of an arraylist       \\
implement floyd warshall                        \\
change seed of random generator                 \\
\hline
\end{tabular}
\end{table}

As a preliminary evaluation of the potential of NLP2Code, we asked ten undergraduate students at the University of Adelaide who are not connected to this research project to use the plugin as part of the development for their coursework for two weeks. In addition to the tasks that they used to invoke NLP2Code, the plugin recorded whether their task came from content assist, how many code snippets they cycled through, and whether they considered the final code snippet to be helpful. For the latter, we asked them at the end of each invocation (i.e., at the first \texttt{Enter} keypress after an invocation) \textit{``Was this code snippet helpful?''}. In addition, we inserted all of their queries into the state-of-the-art tool T2API, a statistical machine translation tool that takes a given English description of a programming task as a query and synthesizes an API usage template for this task~\cite{Nguyen16}, and we compared the results to those produced by NLP2Code.

\subsection{Evaluation with student participants}

At the time of our study, all participants were undertaking the course ``Algorithm \& Data Structure Analysis'' at the University of Adelaide in which they were required to use Java for their course assignments. We provided all participants with instructions on how to use NLP2Code and encouraged them to use the plugin as part of their Java development. Six of the undergraduate students used NLP2Code and shared their usage data with us (response rate 60\%). 

The first two columns of Table~\ref{tab:queries} show the number of invocations that resulted in helpful and unhelpful code snippets, respectively. Our six participants used NLP2Code for 101 invocations, the majority of which resulted in a helpful code snippet. We did not filter out any invocations for the analysis. For about half of the invocations, the content came from content assist. Table~\ref{tab:examples-auto-complete} shows examples of invocations using content assist, and Table~\ref{tab:examples-without} shows invocations without using content assist, i.e., invocations made by highlighting text and pressing \texttt{Ctrl+6} or by surrounding text with question marks. Note that even the invocations without the content assist feature usually follow a task pattern, i.e., starting with a verb followed by a noun phrase, such as \textit{``initialize two dimensional vector''}. These invocations suggest that our underlying assumption that developers need support for tasks is warranted. Our future work will investigate how we can expand NLP2Code's content assist feature to support more of these tasks.

Figure~\ref{fig:cycles} shows the number of ``next snippet'' calls that our participants made using \texttt{Ctrl+`} and how many of these invocations led to a helpful result. In many of the cases, the participants did not cycle through results, but chose the first code snippet returned by NLP2Code. The majority of these snippets was seen as helpful. Other invocations resulted in up to ten ``next snippet'' calls, with similar success rates. We conclude that often, the first code snippet returned by NLP2Code can solve the user's task, and in cases where this does not happen, the ``next snippet'' feature has a high chance of helping the user.

\subsection{Comparison with state of the art}

NLP2Code is most similar to the recently published T2API~\cite{Nguyen16}. In contrast to NLP2Code, T2API has no content assist, its integration into the IDE is not seamless since queries are made in a separate window, and the authors have not evaluated the tool with users. As the name suggests, T2API is specific to API calls whereas NLP2Code can provide code snippets for any task documented in the answers on Stack Overflow. For example, our preliminary evaluation suggests that providing help with algorithms is one of the major benefits of NLP2Code. 

When we inserted the 101 queries that our participants produced into T2API,\footnote{\url{http://atwood.encs.concordia.ca:5905/T2APIRESTService/}} only 16 (i.e., 16\%) resulted in reasonable code snippets, see the last column of Table~\ref{tab:queries}. 40 queries did not return any code snippet, and the remaining 45 resulted in obviously incorrect code snippets. For example, the query \textit{``convert inputstream to string?''} resulted in the following code snippet from T2API: \texttt{String.split(); String.length(); StringBuilder.toString(); String.substring();}, whereas NLP2Code produced \texttt{String myString = IOUtils.toString (myInputStream, "UTF-8");}.\footnote{\url{http://stackoverflow.com/a/350723}} While these findings might not generalize to other settings (e.g., outside of the university environment), we conclude that at least for the kind of queries produced by our participants, NLP2Code outperforms T2API. This is likely explained by the fact that T2API's focus is on API suggestions, not code suggestions.

\section{Related Work}

Similar tools to T2API that also do not provide content assist include the Bing Developer Assistant~\cite{Zhang16}, \textsc{DeepAPI}~\cite{Gu16}, and \textsc{swim}~\cite{Raghothaman16}.

Other prominent tools that integrate Stack Overflow content into the IDE are Seahawk~\cite{Ponzanelli13} and Prompter~\cite{Ponzanelli14}. Seahawk formulates queries automatically based on the active context of an IDE, presents a ranked list of results, and lets users import code snippets through drag and drop. Prompter takes this idea a step further by notifying developers about the available help. Similar to T2API, these two tools utilize additional windows or views in the IDE and do not integrate into a source code editor in the same way that NLP2Code does. In addition, Seahawk and Prompter do not allow users to search for code snippets for a given task.

In the area of code snippet search, Zagalsky et al.~\cite{Zagalsky12} presented Example Overflow, an online code search and recommendation tool. Unlike NLP2Code, Example Overflow is not integrated into an IDE, but is its own website. In their work on making sense of online code snippets, Subramanian and Holmes~\cite{Subramanian13} extracted structural information from short plain-text snippets on Stack Overflow and showed that these structural relationships could improve code snippet search.

\section{Conclusion and Future Work}

We introduced NLP2Code, a content assist for code snippets integrated into the Eclipse IDE. The goal of NLP2Code is to eliminate the context switch that occurs when developers look up code snippets in a web browser while developing in an IDE. NLP2Code closes the vocabulary gap between developers' needs and code snippet meta data through a content assist feature based on natural language tasks extracted from the titles of questions on Stack Overflow. In our preliminary evaluation, NLP2Code returned helpful code snippets in the majority of cases while supporting a wide range of tasks, including API usage examples, template code, and algorithms. 

Based on these encouraging results, our future work includes further improvements to NLP2Code's content assist feature and to the mapping of tasks to code snippets (e.g., avoiding outdated code snippets, selecting those that integrate easily into the existing code base, selecting those that are easy to understand~\cite{Treude17}, or adding additional documentation on-demand~\cite{Robillard17}), as well as the automated integration of code snippets into the existing code base (e.g., by building on Jigsaw~\cite{Cottrell08}). In addition, we plan to evaluate the tool with professional developers in an industry setting.

\section*{Acknowledgements}

We thank the students who participated in our evaluation for using NLP2Code as part of their studies.

\end{sloppy}
\end{document}